\documentclass[11pt,a4paper]{article}
\pdfoutput=1
\usepackage{extracommands}
\usepackage{authblk}
\usepackage{pifont}

\begin{document}

\title{Machine-Readable Privacy Certificates for Services\footnote{A revised
version of this manuscript will appear
in the Proceedings of the International Conference on Secure Virtual Infrastructures
(DOA Trusted Cloud '13) to be held on 9-10 September 2013, in Graz, Austria. If you
wish to refer to this work, please cite~\cite{doa2013} instead.}}

\author[\ding{72}]{Marco Anisetti}
\author[\ding{72}]{Claudio A. Ardagna}
\author[\ding{73}]{Michele Bezzi}
\author[\ding{72}]{\\ Ernesto Damiani}
\author[\ding{73}]{Antonino Sabetta}

\affil[\ding{72}]{Universit\`a degli Studi di Milano, Dipartimento di Informatica, Italy
{\newline  \small e-mail: firstname.lastname@unimi.it}}
\affil[\ding{73}]{SAP Product Security Research, Sophia-Antipolis, France
{\newline  \small e-mail: firstname.lastname@sap.com}}

\date{15 February 2013}

\maketitle
 
\thispagestyle{empty}

\hrule
\begin{abstract}
Privacy-aware processing of personal data on the web of services requires managing a number of issues arising both from the technical and the legal domain.
Several approaches have been proposed to matching privacy requirements (on the clients side)
and privacy guarantees (on the service provider side). Still, the assurance of effective data protection
(when possible) relies on substantial human effort and exposes organizations
to significant  (non-)compliance risks. In this paper we put forward the idea
that a privacy certification scheme producing and managing machine-readable
artifacts in the form of privacy certificates can play an important role towards
the solution of this problem. Digital privacy certificates represent the reasons
why a privacy property holds for a service and describe the privacy measures
supporting it. Also, privacy certificates can be used to automatically select
services whose certificates match the client policies (privacy requirements).

Our proposal relies on an evolution of the conceptual model developed in the \afs project
and on a certificate format specifically tailored to represent privacy properties.
To validate our approach, we present a worked-out instance showing how privacy
property \emph{Retention-based unlinkability} can be certified for a banking financial service.
\end{abstract}

\hrule

\newpage

\setcounter{footnote}{0}
\section{Introduction}
\label{sec:intro}

The success of the Web as a platform for the provisioning of services and the huge amount of personal information disseminated, collected, and managed through the network comes with important concerns for the privacy of users' data.
The growing awareness of users, on one hand, and the increasing regulatory pressure
coming from governments, on the other, is imposing new requirements and constraints
on business that need to handle sensitive data to carry out their services.

As part of the effort of ensuring compliance to the new data protection laws and regulations,
several solutions have been proposed defining different privacy-aware
languages that help users in defining which of their data can be used, by whom, when,
and for which purposes~\cite{ACKLNPSSV.JCS2009,jcs2008,adfps-passat2010,ahks02,xacml,p3p}. 
These solutions constitute an important step towards increasing the availability of
practical data protection technology, leveraging automated processing of privacy policies.
To this end, by capturing the data protection requirements as explicit policies,
they do address a key facet of the problem.
However, their applicability remains limited unless the data protection \emph{guarantees} offered by
providers are expressed similarly in a format that can be processed automatically.
Initiatives such as \emph{EuroPriSe}~\cite{Europrise2008} and \emph{Trust-E}~\cite{trust-e} represent an initial
move in the direction of explicitly representing the data protection measures put in place
by a service (or by a software product in general), but they have significant limitations.
Firstly, they take an \emph{all-or-nothing} approach to compliance (a product is either compliant
with their certification schema or not), which does not allow reasoning about privacy assurance
based on richer, finer-grained client requirements. Secondly, these schemes rely on a format
that, although structured, is essentially based on a natural language description 
meant for human consumption, but that is not suitable for automated processing. Finally,
the evaluation process followed to assess the correctness and adequacy of the
privacy protection measures declared in the certificates is not described in detail,
which makes the evaluation itself somewhat opaque to the client.

Unfortunately, even the adoption of such first-generation privacy certification schemes is quite
an exception; most frequently, service providers release just a text document (typically
a web page in their website) describing what data they handle and what protection measures
they put in place to protect those data. Such a description, expressed in natural language,
requires an understanding of the data protection problem and solution spaces that users
cannot be expected to have. This means that it is extremely difficult for users to determine
if the protection measures offered by a service adequately cover their needs.
Furthermore, the privacy statements associated to services are usually self-declarations
by the service provider, which are difficult (if not impossible) for the client to check.

To address these problems, we believe that \emph{i)}~the privacy protection statements should be expressed
in an explicit, machine-readable format so that the matching of privacy measures (offered by candidate services)
with the corresponding client policies (privacy requirements) can be automated; and \emph{ii)}~the statements by
service providers should be checked and endorsed by a third party that users trust (e.g., a recognized certification entity).
Addressing these two aspects would unlock new scenarios that are not possible today:
users could discover services based on their data protection guarantees, 
determine whether a service fulfills a particular privacy policy, and
compare similar services based on the extent to which each of them targets a
specific data protection goal.

Machine-readable certification of \emph{privacy} statements serve the interests of both clients and providers.
As for clients, explicit privacy certificates mean improved \emph{transparency}.
Companies (especially SMEs) may have not the adequate resources to assess the
``quality'' of offered services, especially from the point of view of security
an privacy. The opportunity of having security and privacy features
described in a structured and machine-readable artifact, as in a digital security
an privacy certificate, can support users to make meaningful comparisons, which
may also be (partly) automated and supported by tools, similarly to what is described
in~\cite{midhat-csp}. 
On the other hand, privacy certificates are an
effective means for service providers to demonstrate compliance with
data protection regulations and customer requirements. 
Although legal compliance with privacy an data protection
regulations are mandatory nowadays, organizations often struggle to deal with the
large diversity of regulation across geographies and sectors. For example, EU data
protection directives often differ from US privacy regulatory framework, not
to mention that the EU directive can be differently implemented in the 27 EU
member states or sector specific regulations (e.g., HIPAA). Privacy certifications
can provide a ``stamp of approval'' of a trusted, expert third-party attesting the
adherence to specific legal and privacy frameworks, ultimately supporting the
users to adopt service-based solutions that are provably compliant to national
and sector specific regulations.

Over the last three years, the \afs project~\cite{assert4soa} has investigated
ways of realizing a novel, light-weight approach to security certification of services,
according to which finer-grained security properties of applications and services are
evaluated by independent third parties and can be expressed in machine-readable artifacts (called \assert{}s).
Among other results, the project defined a conceptual model and a certificate representation, which
provides a concrete structure to represent security certification artifacts. Also,
a reference architecture has been defined to support the processing of certificates,
the discovery of services based on their security properties, and the automated matching
of client requirements with services having corresponding certified properties.

In this paper we present an evolution of the \afs conceptual model and certificate format
that is specifically tailored to represent privacy properties.

The remainder of this paper is organized as follows. Section~\ref{sec:motivex} illustrates
the motivation of our work and a reference scenario. Section~\ref{sec:passert} presents
our certification model for privacy, and Section~\ref{sec:privacy-retention} illustrates its
application using concrete examples of certificates and focusing on privacy property retention-based
unlinkability. Section~\ref{sec:relwork} discusses related work and, finally, Section~\ref{sec:concl} concludes the paper.

\section{Motivating Example}
\label{sec:motivex}

We consider a scenario in which a client is searching for a privacy-aware
IFX-based\footnote{Interactive Financial eXchange (IFX) Standard (\url{http://www.ifxforum.org/standards/}),
a financial messaging protocol initially defined in the 1997 by financial industry and technology leaders. IFX aims to exchange data electronically to accomplish a variety of transactions between (unknown and) distributed entities. The IFX standard supports many financial and security functionalities, and integrates them in a service-oriented architecture.} financial service that addresses its privacy policies (e.g., any personal information provided to the service stays confidential or is deleted after a given period of time). This scenario involves \emph{i)} a client accessing the IFX-based financial service with a set of privacy requirements, \emph{ii)} a service provider implementing an IFX-based service exposed as a SOAP-based service on the Web, and \emph{iii)} a certification authority certifying the privacy properties of web services. 

In particular, we consider an IFX-based service implementing a \emph{Deposit and  Withdrawal} service that enables clients to make deposits (operation \emph{CreditAdd}), possibly via cheque, and withdrawals (operation \emph{DebitAdd}) in/from their bank account, using a reverse ATM. 
This service puts strong requirements on security and privacy of the clients, such as, confidentiality of the messaging exchange, integrity of data, authenticity of the involved parties, privacy of data in the cheques, and introduces the need of a security- and privacy-oriented certification scheme.

In this paper, we focus on the certification of privacy property \emph{retention-based unlinkability}, meaning that the service is certified to maintain the client's personal data following the client's requirements specified through a retention-based privacy policy.
Let us consider the scenario in which a client deposit a cheque using the reverse ATM connected to our \emph{Deposit and  Withdrawal} service. The reverse ATM scans the cheque and allows the client to specify a retention period for the cheque scan when stored at the \emph{Deposit and  Withdrawal} service storage.\footnote{The cheque scan can provide additional information of interest like the cheque transfers and bounces, signatures, dates, which may be sensitive from a privacy point of view.} We note that if the retention period is not specified a default one will be used by the service provider.
The cheque scan is sent as a parameter of the request to operation \emph{CreditAdd} of the \emph{Deposit and  Withdrawal} service, via the SOAP with attachment standard (\url{http://www.w3.org/TR/soap12-af/}), and the retention period specified by the client is associated as an annotation to it. Additional parameters of operation \emph{CreditAdd} (e.g., amount and identity token) are associated with a default retention period possibly dictated by the legislation of the country in which the reverse ATM resides. %

\section{Privacy-Assert}
\label{sec:passert}
We propose the concept of digital privacy certificate for services (\passert). \passert{}s are machine-readable, signed statements, bound to services, that certify the privacy properties guaranteed by a service. As in current certification schemes, the assessment of the property is performed by an independent third party (e.g., a certification authority), who issues (and signs) the \passert.
The certification is based on an evaluation of the service characteristics (e.g., using formal methods or testing), which can be represented in the \passert.  
Differently from existing schemes (e.g., EuroPriSe~\cite{Europrise2008} or Common Criteria~\cite{cc}), \passert{}s are represented as (signed) XML documents, a format suitable for automated reasoning and processing.

\begin{figure}[t]
\includegraphics[width=0.95\textwidth]{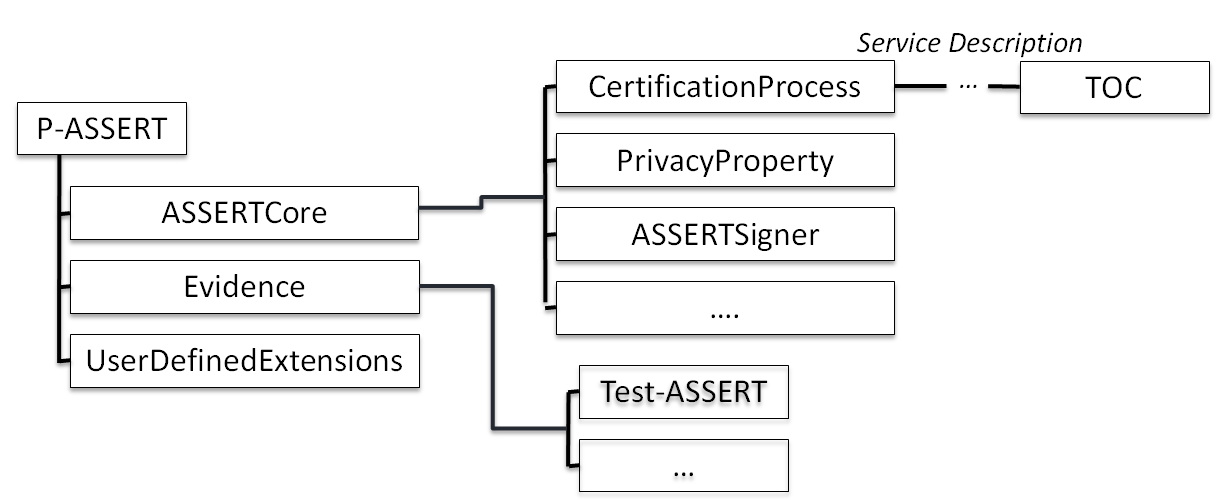}
\caption{\passert: high-level structure.}
\label{fig:core}
\end{figure}

In the following, we present the structure and main features of \passert{}s, which
extends the digital security certificates (\assert) introduced in~\cite{bezzi2011architecture}. As in the original \assert,  each \passert includes two main parts (see Figure~\ref{fig:core}).

\begin{itemize}
\item A {\it Core} part that provides information about the certified entity (service description), the specification of the privacy property of the certified entity, and additional contextual information such as the certification authority, signature, and certification process. 
\item An {\it Evidence} part that describes the details of the evaluation performed by the certification authority (see Section~\ref{sec:evidence}) and supports the certification claims, such as a structured description of the test suites executed as part of the evaluation process (see~\cite{Kaluvuri2013} for details).
\item An additional part reserved for extensions, that can be used, e.g., to cover domain-specific concerns or to provide additional information on top of the content of the \emph{Core} and/or the \emph{Evidence}.
 \end{itemize}

More in detail, in the core part, the service description contains   the Target of Certification (TOC) describing the service being certified, and the Target of Evaluation (TOE) describing the part of the Target of Certification that is evaluated and the rationale for protecting the assets that are identified. More sophisticated models can be present in the {\it evidence} section of the \passert to describe the evaluation performed, e.g., Symbolic Transition System model for the generation of test cases (see Section~\ref{sec:evidence}).

Note that, traditional security (as Common Criteria) or privacy (as EuroPrise) certification schemes do not make a clear distinction  between the system that is being certified
and the aspects of the system that are subject to evaluation, limiting the description to the TOE in  natural language. However, this distinction becomes more relevant whenever we want to use the certificates in service-based systems, because services can be easily composed of multiple, external services, and it should be clear which part is evaluated.
Similarly, to allow for machine-readability, the service description  also provides a list of assets, which will be explicitly referred in the different parts of the certificates. These assets replace the natural language description of assets-to-be protected in today certification schemes. 

The privacy property specification element, contains a multi-level description of the privacy  property at different abstraction levels. We discuss this element in detail in Section~\ref{sec:p_property}.
The evaluation specific portion of the certificate defines the representation of the details and results of the service evaluation process needed to support the certified  property, describing, for example, the models used to generate the test cases and the tests performed on the system. We will describe this part in Section~\ref{sec:evidence}.

The proposed structure of the \passert is based on the the \assert model (which targets security properties), the main difference is in the description of the property. In the next sub-section, we will present the privacy property specification element, we refer the reader to~\cite{bezzi2011architecture,Kaluvuri2013}, for a complete analysis of the remaining part of the \assert .

\begin{figure}[t]
\includegraphics[width=0.95\textwidth]{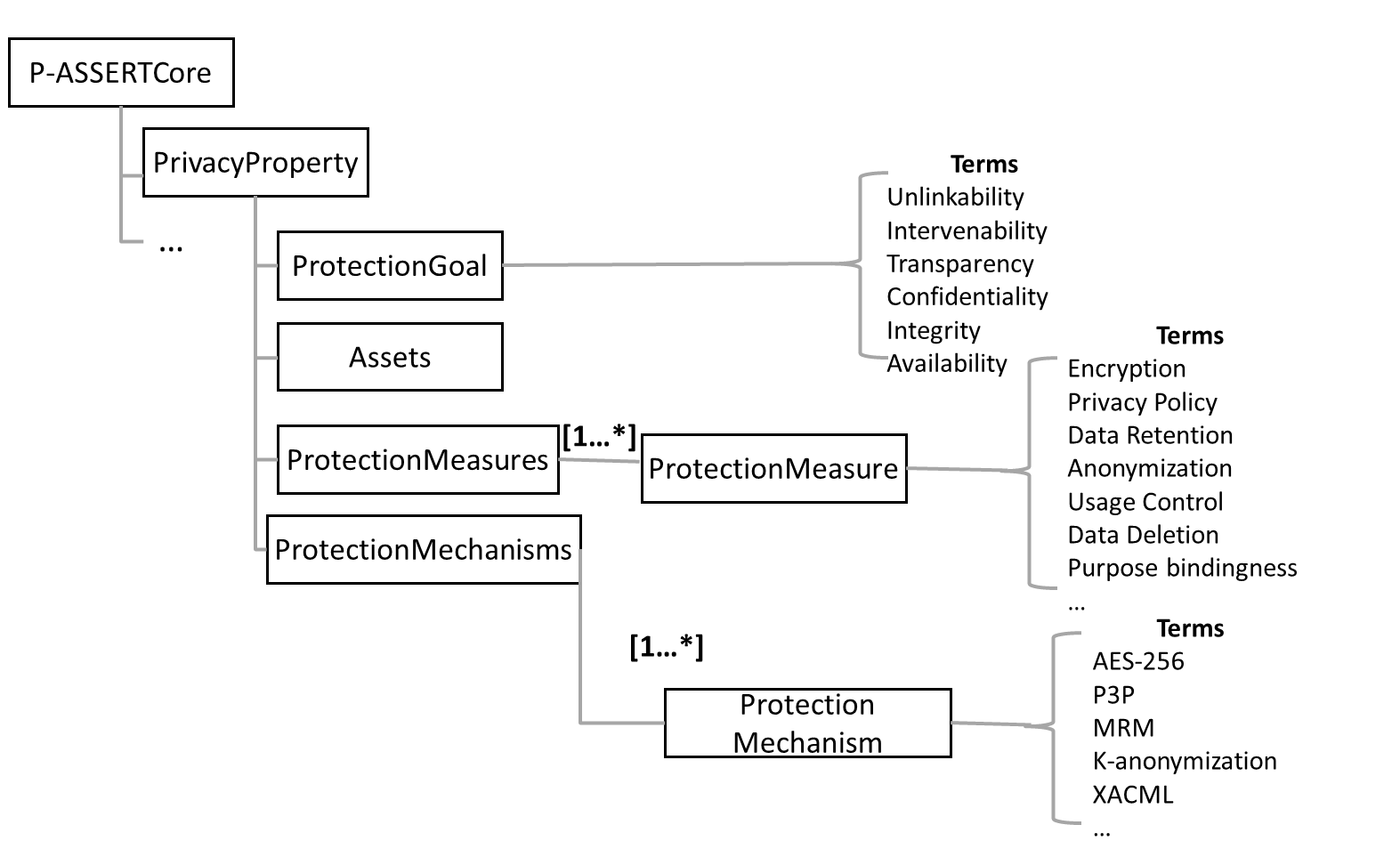}
\caption{{\it PrivacyProperty}: main elements. }
\label{fig:p_property}
\end{figure}

\subsection{Privacy Property}
\label{sec:p_property}

Privacy properties need to be specified at different levels of abstraction, from more abstract concepts to fine-grained representations. The advantage is two-fold: first, it allows end-users with different levels of expertise to understand the privacy features of the service, increasing transparency; second, it permits users (being machines or human beings) to search for services that match their privacy requirements at different levels of complexity. 
Accordingly, we propose different elements, from abstract concepts to the technical implementation mechanisms, to describe the property,
as illustrated in  Figure~\ref{fig:p_property}. A privacy property for a certain asset
({\it Assets} element to the service description) is described with a three layer structure, as in~\cite{Bock2010,Hansen2012},
in terms of {\it Protection Goal}, {\it Protection Measures} and {\it Protection Mechanisms}.
All the other elements are common with the previously introduced \assert representation for security
certificate (see~\cite{Kaluvuri2013} for details).

Regarding the most abstract layer, Privacy by Design principles~\cite{6387956} constitute a natural starting point to express privacy principles. On the other hand, they mix regulative and engineering criteria, making them unsuitable for describing technical features only.
Recently, in analogy with the ``classical'' data security protection goals of {\it confidentiality}, {\it integrity}, and {\it availability}, three additional {\it data protection goals} (and corresponding protection measures) have been proposed~\cite{Bock2010,Hansen2012}.
\begin{description}
\item{\bf Transparency} means that ``the collection and processing
operations of data and its use can be planned, reproduced, checked and evaluated
with reasonable efforts.'' It can be supported by measures such a clear privacy policy, breach notification, and so on.
\item{\bf Unlinkability}  ensures that personal data cannot be linked across  domains or
used for a different purpose than originally intended~\cite{Hansen2012}. 
It can be supported by measures like: limited retention period, data erasure,  anonymization, data minimization, separation of contexts by different identifiers. 
\item{\bf Intervenability} is the ``ability to intervene'' for data subjects, operators and supervisory data protection authorities to apply corrective measures if necessary. For example, it includes the right to rectification and deletion of data for the  data subject.  It can be supported by measures such as mechanisms for handling data subject's correction requests, \emph{break-the-glass} policies, and the like.
\end{description}

These protection goals provide a high-level description of the privacy properties, but they do  not give any information how these goals can be achieved. The \passert contains a list of  {\it protection measures}, which are linked to a protection goal, indicating  the necessary measures  to reach the  goal.
For example, the {\it unlinkability} protection goal can be supported by {\it anonymization} and {\it data retention} measures.
Measures are realized by specific protection mechanisms, describing the techniques or procedures used to realize a specific protection measure. For example, {\it anonymization} protection measure can be implemented by specific {\it k-anonymity} algorithms, with a set value of $k$.

More formally, a privacy property $\pij{}$ is a pair ($\hat{\pij{}}$,$A$), where $\pij{}.\hat{\pij{}}$ is an protection goal and $\pij{}.A$ is a set of class attributes referring to specific characteristics of the privacy function implemented by the service (i.e., detailed description of protection measures and mechanisms).
For instance, property $\pij{}$$=$$($\emph{confidentiality},\{{\tt mea\-sure}$=$encryption,{\tt algo}$=$DES,{\tt key}$=$112bit,{\tt ctx}$=$in transit\}) describes a privacy property whose \textit{protection goal} is confidentiality in transit, \textit{protection measure} is encryption, and \textit{protection mechanism} is DES encryption algorithm with key length of 112bits.

In some cases, a partial order can be defined over privacy properties based on attribute values, inducing a hierarchy \Hie\ of properties as a pair (\T,\domina), where \T\ is the set of properties and \domina\ the partial order. Given two properties \pij{i} and \pij{j}, we write \pij{i}\domina\pij{j}, if \pij{i} is weaker than \pij{j} (see~\cite{AADS.TWEB2013} for a more detailed discussion on partial ordering of security properties).
The hierarchy of privacy properties is fundamental for comparing different services from a privacy point of view, which is one of the most prominent functionality for a privacy-aware SOA infrastructure.

\subsection{Evidence Representation in \passert}
\label{sec:evidence}

In this section, we describe a test-based certification scheme, that is, a process producing evidence-based proofs that a (white- and/or black-box) test carried out on the software has given a certain result, which in turn shows that a given high-level security property holds for that software~\cite{book}. The evidence in \passert contains test-based artifacts and details on how these artifacts support privacy property \pij{} defined in the core part of \passert. More in detail, it is divided into two main sections as follows.

\begin{figure}[!t]
\begin{center}
\SelectTips{cm}{}
\xymatrix @-1pc {
&*++[o][F-]{1} \ar@{->}[dd]^{\scriptsize{\txt{?Signon$<$usr,pwd$>$\\[(usr,pwd)$\neq$null]}}}&\\
~\\
&*++[o][F-]{2} \ar@{->}[ldd]_{\scriptsize{\txt{!Signon$<$result$>$\\[result=failure]}}}\ar@{->}[rdd]^{\scriptsize{\txt{!Signon$<$result,token$>$\\[result=ok]}}}&\\
~\\
*++[o][F-]{3} &&*++[o][F-]{4} \ar@{->}[ddl]_{{\scriptsize{\txt{?CreditAdd$<$amount,token,scan,rp$>$\\[amount$>$0~$\wedge$~token$\neq$null~$\wedge$~scan$\neq$null]}}}}\ar@{->}[ddr]^{{\scriptsize{\txt{?DebitAdd$<$amount,token$>$\\[amount$>$0~$\wedge$~token$\neq$null]}}}}\\
~\\
&*++[o][F-]{5} \ar@{->}[dd]_{\scriptsize{\txt{!CreditAdd$<$result$>$}}}&&*++[o][F-]{7} \ar@{->}[dd]^{\scriptsize{\txt{!DebitAdd$<$result$>$}}}\\
~\\
&*++[o][F-]{6} &&*++[o][F-]{8} \\
}
\caption{\label{fig:sts}WSCL-based model for deposit and withdrawal service.} 
\end{center}
\end{figure}
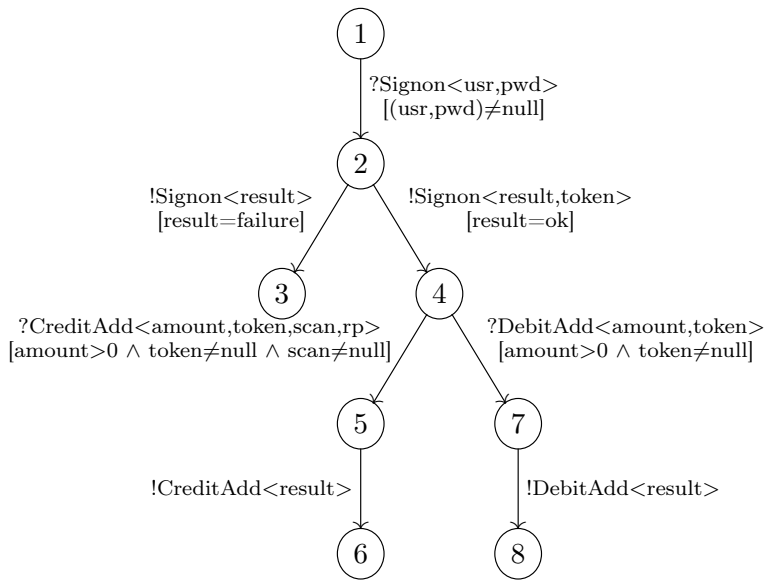

\paragraph{Service model \m:} A Symbolic Transition System (STS)~\cite{tretmans06} that specifies service behavior and interactions as a finite state automaton. It is used for automatic generation of test cases. The service model specifies a label \emph{Model} that describes its level of detail and assumes values in: \emph{i) WSDL} when \m\ models the Web Service Definition Language (WSDL) interface only, \emph{ii) WSCL} when \m\ models the client-server conversation in the Web Services Conversation Language (WSCL) document, and \emph{iii) implementation} when \m\ models the implementation of service operations.
We note that the service model only describes those operations, called Most Important Operations (MIOs), that are needed to certify privacy property \pij{} and to maintain the correctness of the service model.
A set of quantitative indexes \indexes{} (e.g., number of states) %
is defined to calculate a quality measure for the model (the complete set of indexes is provided in~\cite{AADS.TWEB2013}). 
As an example, a WSCL-based model for the \emph{Deposit and  Withdrawal} service described in Section \ref{sec:motivex} is depicted in Figure~\ref{fig:sts}.

Figure~\ref{fig:sts} shows the WSCL conversation that allow the client to access operations \emph{CreditAdd} and \emph{DebitAdd} of the service. First the client has to log into the system. Operation \emph{Signon} returns the variable \emph{result}$=$`ok' with a \emph{token} in case of successful authentication,
 \emph{result}$=$`failure' otherwise. 
After a successful \emph{Signon}, the client can call either operation \emph{CreditAdd} or \emph{DebitAdd}. \emph{CreditAdd} takes as input variables \emph{amount} (the amount of money to be deposited), \emph{scan} (an optional parameter with a scan of the cheque used to transfer money and passed as an attachment to the SOAP message of the request), \emph{token} (an authentication token returned by operation \emph{Signon}), and \emph{rp} (the retention policy attached to variable scan). \emph{DebitAdd} takes as input variables \emph{amount} (the amount of money to be withdrawn) and \emph{token} only. Both operations return the \emph{result} of the execution as output.

\paragraph{Test Evidence \evid{}:} The testing artifacts proving a property for the certified service. Test Evidence \evid{} is composed of the set of test cases executed on the service, their category (i.e., functionality, robustness, penetration) and type (e.g., random input, equivalence partitioning), and a set of test attributes (e.g., the cardinality of the test set). 
It also specifies a set of test coverage metrics (e.g., branch coverage, path coverage) measuring how much the test set covers the service model, and is complete and exhaustive. 
More formally \evid{}=\{\cat$($\evid{}$)$,\type$($\evid{}$)$,\ta$($\evid{}$)$,\tc$($\evid{}$)$,\tr$($\evid{}$),$\metrics{}\} where \cat$($\evid{}$)$ is the test category, \type$($\evid{}$)$ is the test type and \ta$($\evid{}$)$ are the related attributes; \tc$($\evid{}$)$ are the test cases while \tr$($\evid{}$)$ are the results of their execution. \metrics{j} is the set of test coverage metrics. The complete set of metrics is provided in~\cite{AADS.TWEB2013} and can be used to calculate an aggregated quality metric. 

\vspace{0.5em}
Test evidence can support a variety of privacy-related properties. For instance a certification authority can award a privacy certificate \C{}(\pij{},\m,\evid{}) to a service $s$ with privacy property \C{}.\pij{}=(\emph{Confidentiality},\{{\tt measure}$=$encryption,$\mathtt{algo}$=\emph{DES},\\$\mathtt{key}$=\emph{112},$\mathtt{ctx}$=\emph{in transit}\}$)$, service model \C{}.\m=\{\emph{WSCL},$*$\}, and evidence \C{}.\evid{}=\\$($\emph{Functionality},\emph{Input Partitioning.Equivalence Partitioning},\{$\mathtt{card}$=130\},$*$,$*$,$*$$)$,\newline where \C{}.\evid{}.$\mathtt{card}$ is the cardinality of the test set and $*$ means any value. Certificate \C{} provides the evidence proving that $s$ holds the privacy property with the protection goal of confidentiality at communication (\emph{in transit}) level, using a 112-bit \emph{DES} algorithm. In this example, evidence is produced using a WSCL-based model and a set of 130 test cases with test category \emph{Functionality} and test type \emph{Input Partitioning.Equivalence Partitioning}.

\section{Certification of Retention-Based Unlinkability}\label{sec:privacy-retention}

In the previous section, we briefly presented an example of \passert certificate for Confidentiality of data in transit supported by functionality test. In this section, we present a complete worked-out example showing how privacy property \emph{Retention-based unlinkability} can be certified. We assume that a simplified language is available permitting to specify the \emph{retention period} of data contained in a service request.\footnote{This assumption is not restrictive since our solution can be easily adapted for working with any privacy policy language supporting retention, including classic P3P~\cite{p3p}.} %
After the retention period expires, the service provider must delete any reference to the data, assuring users that their data are no longer available for access.

To show the process of certifying a service for property \emph{Retention-based unlinkability}, we consider the \emph{Deposit and  Withdrawal} service in Section~\ref{sec:motivex} implementing a privacy retention mechanism similar to the one adopted by Microsoft Exchange Server 2010, which supports the definition and enforcement of retention tags and policies~\cite{exchange}, but at filesystem level. 
We assume that each request performed by a client specifies a \emph{sticky policy}~\cite{ACKLNPSSV.JCS2009} for each data item with the retention period. Sticky policies are data handling policies attached to the personal data they protect and regulate how personal data will be handled at the receiving parties (i.e., data controllers and processors). Users specify these policies to define restrictions on the retention period of their data, thus keeping a level of control on their information also after its release. 
Clearly, the retention period specified in the request by a user must comply with the requirements for retention defined by the service (see Section~\ref{sec:motivex}); if not specified a default retention period applies for the user request. 

In the following, we consider the certification of a cheque-based \emph{CreditAdd} only and assume that the retention period for a cheque scan attached to the request in a cheque-based deposit is directly specified by the client using the reverse ATM (see parameters \emph{scan} and \emph{rp} in Figure~\ref{fig:sts}).
We note that a retention can be also specified for parameters \emph{amount} and \emph{token}, though not discussed in our scenario. %

Suppose that the service supports a retention mechanism with frequency of control 1 second, minimum retention period 1 day, and maximum retention period 1 year. To this aim, the service implements a mechanism that periodically checks (every 1s) the retention period of each request and deletes all data for which the retention period is expired. In particular, the process implemented by the retention mechanism is composed of the following steps:
\begin{enumerate}
	\item the client sends a \emph{CreditAdd} request to a service annotated with a retention period \emph{rp} for the cheque scan. The retention period is defined in seconds and automatically transformed in a precise date and time at the service side;
	\item the service checks if the retention period complies with its requirements (e.g., minimum retention for cheque scan). If yes, the cheque scan is stored in the filesystem with the retention period; if not, the user request fails;
	\item the service periodically controls the cheques' storage and flags all cheque scan for which the retention period is expired (i.e., the date and time in the retention policy are before the current date and time);
	\item flagged cheques are deleted.
\end{enumerate}

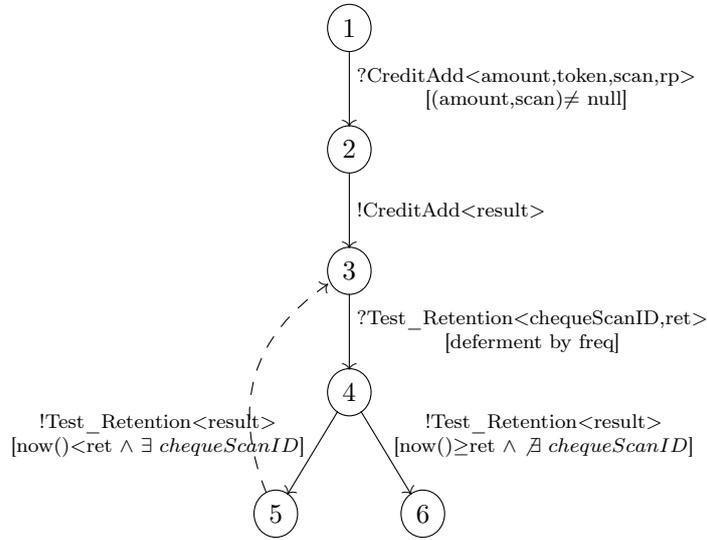
\begin{figure}[!t]
\begin{center}
\resizebox{0.75\textwidth}{!}{
\SelectTips{cm}{}
\xymatrix @-1pc {
&*++[o][F-]{1} \ar@{->}[dd]^{\scriptsize{\txt{?CreditAdd$<$amount,token,scan,rp$>$\\[(amount,scan)$\neq$~null]}}}&\\
~\\
&*++[o][F-]{2} \ar@{->}[dd]^{\scriptsize{\txt{!CreditAdd$<$result$>$}}}&\\
~\\
&*++[o][F-]{3} \ar@{->}[dd]^{\scriptsize{\txt{?Test\_Retention$<$chequeScanID,ret$>$\\[deferment~by~freq]}}}&\\
~\\
&*++[o][F-]{4} \ar@{->}[ldd]_{\scriptsize{\txt{!Test\_Retention$<$result$>$\\[now()$<$ret~$\wedge$~$\exists~chequeScanID$]}}}\ar@{->}[ddr]^{\scriptsize{\txt{!Test\_Retention$<$result$>$\\[now()$\geq$ret~$\wedge$~$\not\exists~chequeScanID$]}}}&\\
~\\
*++[o][F-]{5} \ar@/^2pc/ @{-->} [uuuur]&&*++[o][F-]{6} \\
}

}
\caption{\label{fig:sts2}WSCL-based Test model for Deposit and Withdrawal example.} 
\end{center}
\end{figure}

As soon as the service provider wants to certify this service for privacy property $\pij{}$=(\emph{Unlinkability},\{{{\tt measure}=retention,{{\tt frequency}=1s, {{\tt min\_retention}=1d,\\{{\tt max\_retention}=1y\}), the certification authority (CA) verifies the correct working of the above described retention mechanism by means of \emph{deferred testing execution}.
\emph{Deferred testing execution} is a largely used test execution approach in which the executions of two consecutive test cases are deferred by a specific temporal delay. The delay of the deferment is usually due to the fact that the components that a test case exercises may not be ready for inspection by the time the test runs (e.g., due to instantiation of classes declared with deferred initialization stages). In our approach, execution deferment is used to test properties, like retention, that should become true (or false) after a given period of time.
Our testing strategy relies on the model \m\ of the service (see Figure~\ref{fig:sts2}), on test category \cat, and test type \type\ to produce the test model used for test case generation~\cite{AADS.TWEB2013}. The test model allows deferred test execution via the definition of specific timing conditions on the STS-based service model (see Figure~\ref{fig:sts2}). 
In this specific case, to support the testing activities, the WSDL of the service is extended with a new operation \emph{Test\_Retention}(\emph{chequeScanID},\emph{ret}). This operation, which is used at certification time only and then removed, provides the test code for evaluating the retention mechanism. It takes the name of the file representing the cheque scan (\emph{chequeScanID}) in the service storage and the retention period \emph{ret} (date/time) derived from the retention policy \emph{rp} as input, and returns the result of the testing activity as output. %

The \emph{Test\_Retention} operation checks the storage for cheque \emph{chequeScanID}. The retention mechanism is correctly working and the \emph{Test\_Retention} returns true if: \emph{i)} the cheque is in the storage and the retention period is not expired (while being less than one year) or is expired by less than the 1s (frequency with which the retention is evaluated), \emph{ii)} the cheque is not in the storage and the retention period is expired. It returns false in the other cases.
To test the retention-based privacy property, we first remove operation \emph{Signon} because it is not a MIO and thus does not contribute to the generation of the test cases (see Figure~\ref{fig:sts2}). Also, for simplicity, we removed operation \emph{DebitAdd} and focused on the certification of \emph{CreditAdd}. As a consequence, the test model only involves operations \emph{CreditAdd} and \emph{Test\_Retention}.
Since the retention test is a deferred testing we add deferring time conditions to the STS-based test model in such a way that the \emph{Test\_Retention} can be executed before the retention time is expired ($now()$$<$$ret$) and after the retention time is expired ($now()$$\geq$$ret$).
The test model in Figure~\ref{fig:sts2} generates multiple calls to operation \emph{Test\_Retention} with different deferment times (i.e., \emph{deferment by freq})
for proving the correctness of the retention mechanism implemented by the service. Our model includes a cycle which iterates until the cheque scan is no longer stored by the service, that is,  the retention control mechanism is proved to be correctly implemented for that specific scan. We note that operation \emph{Test\_Retention} is iteratively executed according to the frequency used by the retention mechanism.

Some examples of test cases generated by the test model in Figure~\ref{fig:sts2} are shown in Figure~\ref{fig:testret}.
\begin{figure}[!t]
\scriptsize
\noindent{\sc Privacy Property:} Unlinkability\\
\noindent{\sc Class Attributes:} \emph{measure}=retention, \emph{frequency}=1s, \emph{min\_retention}=1d, \emph{max\_retention}=1y\\
\parbox{5cm}{
\begin{flushleft}
\begin{equation}\label{eq1}
TC1 = \left\{ 
\begin{array}{ll}
I_1:  & DebitAdd(amount,token,scan,rp)\in valid~partitions ~~~~~~~\\ 
EO_1: & result=\emph{ok}\\
PR_2: & deferment~by~freq\\
I_2:  & Test\_Retention(scan.chequeScanID, ret)~with~now()<ret\\ 
EO_2: & result=True\\
\end{array}\nonumber
\right.
\end{equation}

\begin{equation}\label{eq2}
TC2 = \left\{ 
\begin{array}{ll}
I_1:  & DebitAdd(amount,token,scan,rp)\in valid~partitions ~~~~~~~\\ 
EO_1: & result=\emph{ok}\\
PR_2: & deferment~by~freq\\
I_2:  & Test\_Retention(scan.chequeScanID, ret)~with~now()\geq ret\\ 
EO_2: & result=True~\wedge~scan.chequeScanID~is~not~found\\
\end{array}\nonumber
\right.
\end{equation}

\begin{equation}\label{eq3}
TC3 = \left\{ 
\begin{array}{ll}
I_1:  & DebitAdd(amount,token,scan,rp)~with~amount,token,rp\in valid~partitions~\wedge\\
&~~~~~~\wedge~scan\in invalid~partition~~~~~~\\ 
EO_1: & result=\emph{err}\\
PR_2: & deferment~by~freq\\
I_2:  & Test\_Retention(scan.chequeScanID, ret)~with~anytime\\ 
EO_2: & result=True~\wedge~scan.chequeScanID~is~not~found\\
\end{array}\nonumber
\right.
\end{equation}

\end{flushleft}
}
\caption{Test cases for retention for \emph{Deposit and Withdrawal} service.}
\label{fig:testret}
\end{figure}
TC1 and TC2 belong to the functionality test category and consider valid test types (all parameters are in their valid partitions). In general, the test model in Figure~\ref{fig:sts2} will generate a set of TC1-like test cases, until the test time (indicated using \emph{now()}) is greater or equal to the retention time \emph{ret} (in the form of date/time) derived from the user's privacy policy.
TC3 is a robustness test case that verifies whether the cheque scan is invalid (e.g. not correctly scanned), while the other parameters of the \emph{CreditAdd} are valid. In this case the result of \emph{CreditAdd} is an error and the cheque scan must be not saved or deleted immediately from the cheque scan storage; the operation may be maintained in the system log with the other parameters of the function call, depending on the legislation of the country in which the reverse ATM resides. The execution of \emph{Test\_Retention} with a wrong cheque scan must return a \emph{scan not found} independently by the precise time in which it is executed.

The certification authority can also verify the correct support for minimum and maximum retention periods, using the additional test cases showed in Figure~\ref{test}.

\begin{figure}[!t]
\scriptsize
\noindent{\sc Privacy Property:} Unlinkability\\
\noindent{\sc Class Attributes:} \emph{measure}=retention, \emph{frequency}=1s, \emph{min\_retention}=1d, \emph{max\_retention}=1y\\

\parbox{5cm}{
\begin{flushleft}
\begin{equation}\label{eq4}
TC4 = \left\{ 
\begin{array}{ll}
I:  & \emph{request}=CreditAdd~\wedge~(1d<\emph{rp}<1y)\\ 
EO: & result=\emph{ok}\\
\end{array}\nonumber
\right.
\end{equation}

\begin{equation}\label{eq5}
TC5 = \left\{ 
\begin{array}{ll}
I:  & \emph{request}=CreditAdd~\wedge~\emph{rp}<1d\\ 
EO: & result=\emph{error}\\
\end{array}\nonumber
\right.
\end{equation}

\begin{equation}\label{eq6}
TC6 = \left\{ 
\begin{array}{ll}
I:  & \emph{request}=CreditAdd~\wedge~\emph{rp}>1y\\ 
EO: & result=\emph{error}\\
\end{array}\nonumber
\right.
\end{equation}

\end{flushleft}
}
\caption{Test cases for verifying retention period boundary values.}
\label{test}
\end{figure}

\section{Related Work}
\label{sec:relwork}

Different languages and format for machine-readable privacy policy for web applications and services have been proposed.
The XML-based P3P (Platform for Privacy Preferences Project) language
and APPEL (A P3P Preference Exchange Language)~\cite{p3p}
are used for describing privacy policies and privacy negotiations between a web site
and  users. 
The Enterprise Privacy Authorization Language (EPAL)~\cite{EPAL} is based on the same concepts of the eXtensible Access Control Markup Language (XACML)~\cite{xacml}, but 
it implements more privacy-specific conditions, such as purpose-based access control. 
More recently,  the PrimeLife Privacy Language (PPL)~\cite{PPL} was defined as an extension of the XACML authorization language. PPL allows to handle  complex privacy policies, including specifying  secondary usage of the data (e.g., when an external data processor is involved) and privacy obligation handling, relying on XACML for access control conditions.

These privacy policy languages allow for a description and processing of privacy policies, but they do not provide the elements for specifying the protection measures used nor for detailing the evidences supporting their correct implementation.
As mentioned in Section~\ref{sec:privacy-retention}, \passert can rely on policy languages for expressing the privacy conditions, and complement them with a more granular description of the protection measures and evidences.

Another major source of related work for this paper resides in the area of software testing. Similarly to this paper, several works (e.g.,~\cite{Chandramouli2004,Jurjensa2008,Zulkernine2009}) focused on testing non-functional requirements of software systems. 
As far as web service testing is concerned, the line of research closest to the one in this paper considers the problem of testing a web service to assess its correct functioning and to automatically generate test cases used in the verification process~\cite{bozkurt2010,survey}. 
Heckel and Lohmann~\cite{heckel} propose a solution for testing web services that uses Design by Contract and adds behavioral information to the web service specifications. 
More recently, Bentakouk et al.~\cite{bentakouk2011} propose a solution using STS-based testing and STM solver to check the conformance of the composite service implementation with respect to its specifications and/or client requirements. Endo and Simao~\cite{Endo2011} present a model-based testing process for service-oriented applications. 
Existing approaches have mainly focused on static or dynamic testing, while they have not focused on certification. More in detail, these approaches test services with the scope of verifying their security mechanisms (i.e., policy enforcement) ex-post.
The \passert approach elaborates on the approach presented in \cite{AADS.TWEB2013} and concentrates on container-level certification, which implies security policy testing and certification. 
%The certification of privacy property let emerges the importance of the general problem of the certification of security policies. 
In~\cite{Salva2010}, a security testing method for stateful Web Services is proposed. It defines specific (i.e., for each security property to be tested) security rules, eventually derived from policy, using Nomad language with the scope of generating test cases. 
This rule set allows to test different properties such as availability, authorization, and authentication by means of malicious requests based on random parameters, or on SQL and XML injections. The rules are applied to the operation set (obtained from service specifications like WSDL) of the service under test, to generate test requirements (modeled as STSs), which are then synchronized with the specifications to produce the test set.
The goal is to perform a black-box testing of web services exploiting a rule-based approach for the generation of an ad hoc test set. The test set is aimed at discovering if the service under test violates or not the security rules.
%{\bf In our case, we would like to derive a similar set of test rules from the WS-security policy in place and from the gray-box model we adopt for services (e.g. we may be able to have a description of service behavior at interface level with WSCL and at implementation level).}
Other approaches focusing on general testing or on authentication and authorization policies, construct abstract test cases directly from models describing policies~\cite{letraon07,Martin-2006,Mouelhi-2008}. Le Traon et al.~\cite{letraon07} proposed test generation techniques to cover security rules modeled with OrBAC. They identified rules from the policy specification and generated abstract test cases to validate some of them via mutation. 
In \cite{Martin-2006}, the authors developed an approach for random test generation from XACML policies. The policy is analyzed to generate test cases by randomly selecting requests from the set of all possible requests.
In \cite{Mouelhi-2008}, the authors proposed a model-driven approach for testing security policies in Java applications. The policy is modeled with a control language such as OrBAC and translated into XACML.
In our case we describe an approach for testing privacy-specific policies for web services and how is it possible to generate a certification scheme for them.

%##############################################

%
\section{Conclusions}
\label{sec:concl}

We proposed a representation for digital privacy certificates (\passert), which describes the outcome of a privacy certification process for web services in a machine-readable format. These structured and machine-readable certificates enable the service consumer to: \emph{i)} know the details about the privacy features of the service ({\it transparency}), \emph{ii)} access detailed information on the evidence supporting the claim of the certificate ({\it assurance}) \emph{iii)} automatically reason about privacy properties (e.g., allowing service discovery based on privacy requirements). Our proposal extends the security certification framework developed in the context of the European project \afs. Following the same approach, we described a digital privacy certificated supported by  a model-based testing process, which allows to automatically produce evidence that a given privacy property holds for the service. The corresponding machine-readable certificate  contains the certified property, the model of the service used for the automatic generation of the test cases, and the evidence produced by their execution. We also provided a worked-out example of the application of our scheme to the certification of privacy property \emph{retention-based unlinkability}. In this context, we introduced the concept of \emph{deferred testing} as a testing activity that specifies the time intervals between consecutive test cases. Our example focused on \emph{offline} deferred testing, meaning that the test cases on retention policies are executed in an accredited Lab in the framework of a certification process and the retention periods are randomly generated to maximize the coverage of their domain. We note however that our solution can support \emph{online} deferred testing, verifying the correctness of the retention mechanism implemented by the service on real client requests and client-defined retention periods. We plan to further analyze this post-deployment testing scenario in our future work. Our future work will also evaluate the efficiency of our approach, analyzing the overhead required for automatic test generation at the increasing of policy complexity.

\paragraph{Acknowledgements.} This work was partly supported by the EU-funded projects
\afs (contract n. FP7-257351) and \textsc{Cumulus} (contract n. FP7-318580). We thank Kirsten Bock and Marit Hansen for  fruitful discussions.

\bibliographystyle{plain}
\bibliography{biblio}

\end{document}